\documentclass[
twocolumn,
]{ceurart}
\usepackage{algorithmic}
\usepackage{algorithm}
\usepackage{listings}
\begin{document}

\copyrightyear{2024}
\copyrightclause{Copyright for this paper by its authors.
  Use permitted under Creative Commons License Attribution 4.0
  International (CC BY 4.0).}

\conference{RecSys in HR'24: The 4th Workshop on Recommender Systems for Human Resources, in conjunction with the 18th ACM Conference on Recommender Systems, October 14--18, 2024, Bari, Italy.}

\title{Parallel and Mini-Batch Stable Matching for Large-Scale Reciprocal Recommender Systems}


\author[1]{Kento Nakada}[
email=Kento.Nakada@sony.com
]
\cormark[1]
\address[1]{Sony Network Communications, Inc.}

\author[2]{Kazuki Kawamura}[
email=kwmr@acm.org
]
\address[2]{The University of Tokyo}

\author[1]{Ryosuke Furukawa}[
email=Ryosuke.Furukawa@sony.com
]


\begin{abstract}
Reciprocal recommender systems (RRSs) are crucial in online two-sided matching platforms, such as online job or dating markets, as they need to consider the preferences of both sides of the match.
The concentration of recommendations to a subset of users on these platforms undermines their match opportunities and reduces the total number of matches.
To maximize the total number of expected matches among market participants, stable matching theory with transferable utility has been applied to RRSs. 
However, computational complexity and memory efficiency quadratically increase with the number of users, making it difficult to implement stable matching algorithms for several users. 
In this study, we propose novel methods using parallel and mini-batch computations for reciprocal recommendation models to improve the computational time and space efficiency of the optimization process for stable matching.
Experiments on both real and synthetic data confirmed that our stable matching theory-based RRS increased the computation speed and enabled tractable large-scale data processing of up to one million samples with a single graphics processing unit graphics board, without losing the match count.
\end{abstract}


\maketitle

\section{Introduction}

Two-sided online dating platforms, such as those found in job search and online dating markets, have become increasingly popular.
Reciprocal recommender systems (RRSs) are used on these platforms.~\cite{palomares2021reciprocal, Luiz2010RECON}.
In a two-sided matching platform, a recommender system that only considers prediction accuracy may inadvertently cause recommendation inequalities for two main reasons.
First, on a two-sided platform, preferences from both sides of the market determine the success of a match.
Recommendations based solely on the interests of one side are ineffective, and the recommender system should only recommend when both users have mutual interests.
Second, users tend to have difficulties finding compatible partners. 
For example, on job platforms, employers frequently have a limited number of available interview slots because of time constraints. 
If the system recommends the same company to many candidates, it may exceed the employer's capacity, leading to missed opportunities for those candidates who applied but were not granted an interview.
The transferable utility (TU) matching model~\cite{shapley1971tumatching, becker1973} is a framework that considers these matching problems under the assumption that utilities, such as money, can be transferred between matching parties. 
This allows for resource redistribution among market participants, thereby achieving stable matching states.
Choo and Siow~\cite{choo2006who} applied the TU matching market model to RRSs, which demonstrated higher matching chances in empirical matching applications with TUs.
Although stable matching methods have a solid theoretical background, most of them face computational feasibility bottlenecks when executed on real data sizes exceeding 10k. Thus, they are not practically feasible for large-scale user platforms and are only useful for the experimental extraction of a subset of actual service users~\cite{tomita2023fast} or matching at the level of the user group clustered by attributes~\cite{Chen2023, saini2019privatejobmatch}.

\begin{figure}[tbp]
    \includegraphics[width=0.45\textwidth]{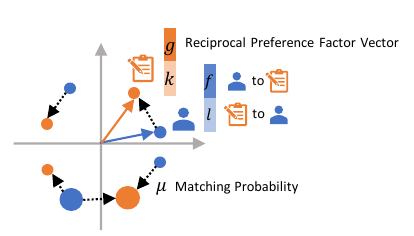}
  \caption{We propose a fast and memory-efficient solution to the TU matching problem, by viewing it as an optimal transport problem with transport costs associated with the inner product of the preference factor vectors.}
\label{method_image_a}
\vspace{-0.5cm}
\end{figure}

In this study, we improve the computational efficiency of the iterative proportional fitting procedure (IPFP), a coordinate descent algorithm used to achieve a stable matching state~\cite{Galichon2021}, through parallelization and online mini-batch computation.
For computational efficiency, we propose a matrix-vector multiplication-based parallel computation method following Sinkhorn's algorithm~\cite{Knopp1967Sinkhorn}, which considers stable matching as an entropy-regularized optimal transport problem. 
To improve memory efficiency, we propose a mini-batch update method assuming a model that uses user factor vectors in the unilateral recommendation model, such as matrix factorization~\cite{Koren2009MF}.
The user set is divided into several partitions (i.e., mini-batches), and a portion of the IPFP coordinate vectors is sequentially updated.
This approach enables efficient estimation of update values when the overall preference scores for user combinations do not fit in memory.

In summary, the following are the main contributions of this study.
\begin{itemize}
    \item We propose an optimization method for RRSs based on the TU matching model to increase computational efficiency using parallel processing.
    \item Furthermore, we propose a memory-efficient mini-batch update method that does not require approximation and works even for large sizes that do not fit in a single memory.
    \item Experiments on a synthetic dataset with up to a million users verify the computational efficiency of the proposed method.
\end{itemize}
To the best of our knowledge, this study is the first to demonstrate how RRSs based on stable matching theory work with large-scale data.

\section{Related Work}
Reciprocal recommender systems (RRSs) aim to achieve matching by considering the preferences of both parties, which are primarily used in areas where mutual matching is crucial, such as dating sites, talent recruitment platforms, and social networking services~\cite{palomares2021reciprocal}.

Existing RRSs first predict unilateral preferences that represent the preference of one user toward another. 
In unidirectional preference calculation, various recommendation methods are employed in recruitment matching problems, including content-based and collaborative filtering approaches ~\cite{7856703}. Among these, collaborative filtering is particularly well suited for scenarios with abundant interaction logs. Herein, we focus on collaborative filtering-based methods for reciprocal recommender systems to apply matching platforms with number of users.

These unilateral preferences are then aggregated to calculate reciprocal preference scores, which serve as recommendation probabilities. 
Aggregation functions that independently calculate for each users pair---such as arithmetic mean~\cite{Luiz2010RECON}, harmonic mean~\cite{Luiz2010RECON, xia2015reciprocal}, and weighted average~\cite{Akiva2018WeightedRRS}---can scale to large data size~\cite{Ramanathan2021ARE}. 
Although these methods can compute for a considerable number of user data, they lack theoretical background and cannot consider constraints on users' limited capacity of matches.

RRSs based on fairness-aware aggregation models redistribute recommendation scores to address the concentration of recommendation issues~\cite{Fedor2017Lijar, do2023optimizing, Su2022}.
Several approaches have attempted to prevent concentration in job recommendation by solving the entropy-regularized optimal transport (OT) problem using reciprocal preference scores as a transport cost matrix~\cite{Guilla2021OT, Yoosof2023Recon}. 
These methods can be applied to large-scale data following Sinkhorn's algorithm~\cite{cuturi2013sinkhorn}. However, there is a lack of theoretical background on the alignment between the OT and equilibrium-matching states in terms of modeling user behavior.

The stable matching theory~\cite{shapley1971tumatching} is a behavioral model based on game theory that balances supply and demand in matching markets.
Galichon and Salani{\'{e}}~\cite{Galichon2021} demonstrated that the TU matching theory corresponds to the dual problem of a special entropy-regularized OT problem, which provides a theoretical foundation for the application of stable matching in recommender systems.

Although stable matching methods have a solid theoretical background, most of them face computational feasibility bottlenecks when applied to real data sizes exceeding tens of thousands.
Galichon and Salani{\'{e}}~\cite{Galichon2021} proposed that their optimization method could be computed in parallel, although their experiment used a dataset of up to a size of 5000.
Chen et al.~\cite{Chen2023} applied stable matching to male--female matching app data of approximately 1000.
Tomita et al.~\cite{tomita2023fast} proposed a TU matching theory-based memory-efficient inference method for reciprocal scores.
To optimize the parameters, all user combinations must be loaded into memory, and a sample size of up to 1000 was used in the experiments.
Thousands to hundreds of thousands of users use job or date matching services. Therefore, we extend the applicability of TU matching to the user bases of that scale.




\section{Proposed Method}

According to~\cite{Chen2023}, this section provides a brief overview of how the TU matching model can be formulated as a convex optimization problem, as described in Section \ref{TU_Matching}.
Thereafter, we propose a parallel update method using a mini-batch approach for the iterative optimization algorithm in Sections \ref{section_Parallel} and \ref{section_Mini-batch}. 
This method enables tractable calculations in near-linear memory consumption, even for many market users.

\subsection{Matching Model with Transferable Utility}
\label{TU_Matching}
We assume matching between candidate $x \in \mathcal{X}$ and employer $y \in \mathcal{Y}$.
The concept of utility explains why an individual chooses where to apply or whom to recruit.
Utilities $U_{x,y}, V_{x,y}$ that candidate $x$ or employer $y$ gains by matching are expressed as follows:
\begin{equation}
\begin{aligned}
U_{x,y} &= p_{x,y} + \epsilon_{x,y}, \: \epsilon_{x,y} \sim \mathcal{P}, \\
V_{x,y} &= q_{y,x} + \eta_{y,x}, \: \eta_{y,x} \sim \mathcal{Q},
\end{aligned}
\end{equation}
where $p_{x,y}$ denotes an observable utility from candidate $x$ to employer $y$, $q_{y,x}$ denotes an observable utility from $y$ to $x$, and $\epsilon_{x,y}$ and $\eta_{y,x}$ are unobserved random utilities with probability distributions of $\mathcal{P}$ and $\mathcal{Q}$, respectively.
We denote the individual who is not matched to any counterpart as 0. 
Let $\mathcal{X}_0=\mathcal{X} \cup\{0\}$ and $\mathcal{Y}_0=\mathcal{Y} \cup\{0\}$ be a set of groups that can be selected as potential partners.




In the TU matching model, we assume that the transferable utility $\tau_{x,y}$ is paid
from an employer to a job candidate upon matching (for example, to adjust supply and demand). 
The matching results are adjusted by optimizing the utility. Specifically, to mitigate the over-concentration of recruitment efforts on highly sought-after candidates, a high cost may be incorporated into the scouting process.
Consider observable joint utility $\phi_{x,y} = p_{x,y} + q_{y,x}$ and probability distribution $\mu_{x,y}$ that specifies a match between candidate $x$ and employer $y$.
Assuming that the distribution of random utilities $\mathcal{P}$ and $\mathcal{Q}$ are independent and identically distributed (i.i.d.) with a type-I extreme value distribution with scale parameter $\beta > 0$, 
Galichon and Salanié~\cite{Galichon2021} revealed that optimizing TU $\tau_{x, y}$ to maximize the expected number of matching in the market is defined as the following convex optimization problem, which maximizes social welfare $W$:

\begin{equation}
\label{primary optimization}
\begin{aligned}
W(\boldsymbol{\Phi})&=\max _{\boldsymbol{\mu} \in \mathbb{R}^{\mathcal{X} \times \mathcal{Y}}}\left(\sum_{x \in \mathcal{X}, y \in \mathcal{Y}} \mu_{x,y} \phi_{x, y}+\beta E(\boldsymbol{\mu})\right), \\
\text { s.t. } \sum_{y \in \mathcal{Y}_0} \mu_{x,y}&=n_x \quad \forall x \in \mathcal{X}, \: \sum_{x \in \mathcal{X}_0} \mu_{x, y}=m_y  \quad \forall y \in \mathcal{Y},
\end{aligned}
\end{equation}
where

\begin{equation}
E(\boldsymbol{\mu})=-\sum_{\substack{x \in \mathcal{X} \\ y \in \mathcal{Y}_0}} \mu_{x,y} \log \frac{\mu_{x,y}}{n_x} -\sum_{\substack{y \in \mathcal{Y} \\ x \in \mathcal{X}_0}} \mu_{x,y} \log \frac{\mu_{x,y}}{m_y}
\end{equation}
is the standard entropy; $n_x$ and $m_y$ are the normalized capacity constraints of candidate group $x$ and employer group $y$, respectively.
\footnote{Decker et al.~\cite{DECKER2013778} investigated the existence and uniqueness of the equilibrium-matching solution.}
In recruitment matching problems, $n_x$ and $m_y$ can correspond to the number of applications a candidate can submit or the number of positions an employer wants to fill. 
The parameter $\beta$ controls the weight of the entropy term in (\ref{primary optimization}). As $\beta$ increases, it promotes a more uniform match result that is less dependent on individual preferences.


Equation (\ref{primary optimization}) is an OT problem with entropy regularization.
We adopt the IPFP proposed in ~\cite{Galichon2021}, a coordinate decent method to solve (\ref{primary optimization}).
The detailed derivation of the TU matching optimization is described in Appendix \ref{Appendix_tu}.

\subsection{Parallel Computation of TU Matching Optimization}
\label{section_Parallel}

%
We propose a parallel update method to alleviate the computational limitation of an IPFP update on a large user base.
At the optimum, an equation on a derivative of (\ref{primary optimization}) derives the following relation between $\phi_{x,y}$ and $\mu_{x,y}$:
\begin{equation}
\label{stable_condition}
\mu_{x,y}=\exp \left(\frac{\phi_{x,y}}{2\beta}\right) \sqrt{\mu_{x,0} \mu_{0,y}}.
\end{equation}

By substituting this expression into the constraints in (\ref{primary optimization}), we obtain the following equation:
\begin{equation}
\label{IPFP}
\begin{aligned}
& \mu_{x,0}+\left(\sum_{y \in \mathcal{Y}} \exp \left(\frac{\phi_{x,y}}{2\beta}\right) \sqrt{\mu_{0,y}}\right) \sqrt{\mu_{x,0}}=n_x, \\
& \mu_{0,y}+\left(\sum_{x \in \mathcal{X}} \exp \left(\frac{\phi_{x,y}}{2\beta}\right) \sqrt{\mu_{x,0}}\right) \sqrt{\mu_{0,y}}=m_y .
\end{aligned}
\end{equation}
The IPFP algorithm solves (\ref{IPFP}) given scaling vector definitions of $u_{x} = \sqrt{\mu_{x,0}}$ and $ v_{y} = \sqrt{\mu_{0,y}},$ using $i$ as the number of iteration steps to repeatedly run through the following updates:
\begin{equation}
\label{iteration}
\left\{\begin{array}{l}
u_{x}^{(i+1)}=\sqrt{n_x+ s_x^2} - s_x  \text { where } s_x= \frac{1}{2} \sum_{y \in \mathcal{Y}} \exp \left(\frac{\phi_{x,y}}{2\beta}\right) v_y^{(i)}  \\
v_{y}^{(i+1)}=\sqrt{m_y+ s_y^2} - s_y  \text { where } s_y= \frac{1}{2} \sum_{x \in \mathcal{X}} \exp \left(\frac{\phi_{x,y}}{2\beta}\right) u_{x}^{(i+1)} 
\end{array},\right.
\end{equation}
Once the algorithm has converged after several iterations, the stable match patterns $\boldsymbol{\mu}$ are calculated according to Equation (\ref{stable_condition}). 

Equation (\ref{iteration}) can be further expressed in matrix-vector arithmetic as follows:
\begin{equation}
\label{iteration_matrix}
\left\{\begin{array}{l}
\boldsymbol{u}^{(i+1)}=\sqrt{\boldsymbol{n}+ \boldsymbol{s}^2} - \boldsymbol{s}  \text { where } \boldsymbol{s}= \frac{1}{2}\boldsymbol{A}\boldsymbol{v}^{(i)}  \\
\boldsymbol{v}^{(i+1)}=\sqrt{\boldsymbol{m}+ \boldsymbol{s}^2} - \boldsymbol{s}  \text { where } \boldsymbol{s}= \frac{1}{2}\boldsymbol{A}^T\boldsymbol{u}^{(i+1)} 
\end{array}\right. ,
\end{equation}
where $\boldsymbol{A}=\exp\left(\frac{\boldsymbol{\Phi}}{2\beta}\right)$.
We denote the update method by Equation (\ref{iteration_matrix}) as \textit{batch} IPFP.
Because Equation (\ref{iteration_matrix}) only involves the matrix-vector product, it can be efficiently computed through parallel computation. However, because the size of the matrices scales in $O(|\mathcal{X}||\mathcal{Y}|)$, the equation is computationally intractable within the available memory. 

\subsection{Mini-batch Computation of TU Matching Optimization}
\label{section_Mini-batch}
To alleviate the memory space limitation, we further propose a memory-efficient update method, \textit{mini-batch} IPFP.
The inner product of the $D$ dimensional factor vectors $\boldsymbol{f}_x, \boldsymbol{g}_y, \boldsymbol{k}_x, \boldsymbol{l}_y \in \mathbb{R}^D$ through matrix factorization is assumed to compute unilateral preferences $p_{x,y}$ and $q_{y,x}$.
\begin{equation}
p_{x,y} = \left\langle\boldsymbol{f}_{x} , \boldsymbol{g}_{y}\right\rangle, \: q_{y,x} = \left\langle\boldsymbol{k}_{x} , \boldsymbol{l}_{y} \right\rangle
\end{equation}
In such a case, user sets can be substituted for the IPFP algorithm to be executed online.
Let $\mathcal{X}_j (j=1,...,J_x, 1 < J_x \leq |\mathcal{X}|)$ and $\mathcal{Y}_j (j=1,...J_y, 1 < J_y \leq |\mathcal{Y}|)$ be a $j$-th mini-batch of the candidate and employer sets, respectively. 
Now, Equation (\ref{iteration_matrix}) can be calculated for each $j$-th mini-batch.

\begin{equation}
\label{iteration_minibatch}
\left\{\begin{array}{l}
\boldsymbol{u}_j^{(i+1)}=\sqrt{\boldsymbol{n}_j+\boldsymbol{s}_j^2}-\boldsymbol{s}_j \text { where } \boldsymbol{s}_j=\frac{1}{2} \boldsymbol{A}_j \boldsymbol{v}^{(i)} \\
\boldsymbol{v}_j^{(i+1)}=\sqrt{\boldsymbol{m}_j+\boldsymbol{s}_j^2}-\boldsymbol{s}_j \text { where } \boldsymbol{s}_j=\frac{1}{2} (\boldsymbol{A}^T)_j \boldsymbol{u}^{(i+1)}
\end{array} \quad ,\right.
\end{equation}
where subscript $\ast_j$ denotes the indices of users in the $j$-th mini-batch and
\begin{equation}
\boldsymbol{A}_j=\exp\left(\frac{\boldsymbol{F}_{j} \boldsymbol{G}^T + \boldsymbol{K}_{j} \boldsymbol{L}^T}{2\beta}\right), \:
(\boldsymbol{A}^T)_j=\exp\left(\frac{\boldsymbol{F} \boldsymbol{G}^T_j + \boldsymbol{K} \boldsymbol{L}^T_{j} }{2\beta}\right).
\end{equation}

According to Tomita et al.~\cite{tomita2023fast}, once optimal $\boldsymbol{u}$ and $\boldsymbol{v}$ are obtained, stable match patterns $\log{\boldsymbol{\mu}}$ can be calculated as the dot product of the following two vectors with dimensions $(2D+2)$.

\begin{equation}
\begin{aligned}
    \log \left(\mu_{x, y}\right) &=\frac{1}{2 \beta}\left\langle\boldsymbol{\psi}_x, \boldsymbol{\xi}_y \right\rangle, \\
    \boldsymbol{\psi}_x &= \text{Concat} \left(\boldsymbol{f}_x, \boldsymbol{k}_x, \beta \log \left(\boldsymbol{u}\right), 1\right), \\
    \boldsymbol{\xi}_y &= \text{Concat}\left(\boldsymbol{g}_y, \boldsymbol{l}_y, 1, \beta \log \left(\boldsymbol{v}\right)\right),
\end{aligned}
\end{equation}
where Concat(*) denotes the concatenation of vectors. 
In the case of large data sizes (i.e., $D \ll |X|$), 
the space complexity is reduced to $O(|X|)$ or $O(|Y|)$. 
In practice, this reduction allows the execution of the IPFP by adjusting the batch size to fit within the memory limit.

The pseudocodes of the \textit{batch} and \textit{mini-batch} IPFP algorithm is shown in Appendix \ref{Appendix_alg}. 

\begin{figure}[tbp]
    \includegraphics[width=0.45\textwidth]{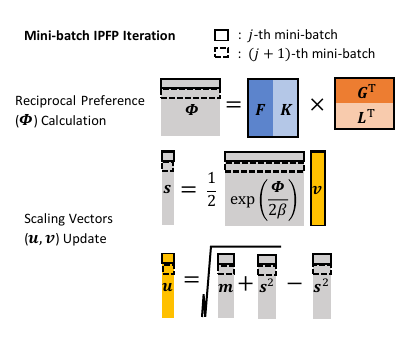}
  \caption{Our Mini-batch IPFP update keeps only a part of the matrices in memory during the update step, achieving parallel computation and memory efficiency.}
\label{method_image_b}
\vspace{-0.3cm}
\end{figure}

\section{Experiments}
Here, 
we first validate the IPFP algorithm by comparing it with existing methods in terms of the expected number of matches, using both real and synthetic data.

Thereafter, as our contribution, we validate the computational efficiency of the proposed batch and mini-batch updates in the CPU/GPU settings. 
Furthermore, we investigate the memory efficiency and calculation performance of the mini-batch IPFP algorithm using various batch sizes.
We aim to improve the efficiency of the IPFP algorithm because the expected number of matches is the same as that of IPFP. 

\subsection{Experiments on Expected Number of Matches}
\label{appendix_sw}
\subsubsection{Datasets.}
We compared the IPFP algorithm with existing methods using data from the online dating platform Libimseti \cite{xia2015reciprocal}. This dataset includes user reciprocal ratings. We chose 500 male and 500 female users who submitted the highest number of ratings. Any missing ratings were filled in using probabilistic matrix factorization with the alternating least squares (ALS) method ~\cite{NIPS2007_d7322ed7}.

In experiments with real data, it is impossible to know the true preferences of each user in the market. Consequently, the expected number of matches can only be calculated using estimated preferences. To address this limitation, we also conducted experiments using synthetic data, which allowed us to control the true preferences in a structured manner.

We evaluated the methods using synthetic data generated based on the process described in \cite{Su2022}. 
The preference matrices $p_{x,y}$ and $q_{y,x}$ are generated by interpolating random values sampled from an independent uniform distribution and values proportional to the index of samples, indicating crowding of preferences in the market. The degree of preference crowding can be adjusted using the parameter $\lambda \in [0, 1]$.
The experiment involved setting the number of employers at 500, candidates at 1000, and the crowding parameter was varied over ${0.0, 0.25, 0.5, 0.75}$.
The observational data are simulated by sampling binary values $\{0,1\}$ from Bernoulli distributions based on the probabilities of the generated preference matrices.
The factor vectors are obtained by the implicit alternating least squares (iALS) method \cite{Paterek2007ALS} from observational data.

For both real and synthetic dataset, we repeated this evaluation 10 times to obtain the average and standard error.

\subsubsection{Algorithms and Metrics.}
We compared the TU method with three baselines: naive, reciprocal and cross-ratio (CR) methods. 
The naive method uses the unidirectional preference from the candidate to employer $p_{x,y}$ to create the presentation list.
The reciprocal method uses the product of preferences from both sides $p_{x,y} * q_{y,x}$ to create the presentation list.
The CR method proposed in \cite{James19LFRR} uses a cross ratio uninorm of preferences from both sides.
We attempted to apply the method proposed by Su et al.\cite{Su2022} to both real and synthetic data, however the optimization process did not complete in tractable time. 

We further compared the optimization methods of the TU method: batch and mini-batch IPFP. 
For the TU method, we used $\beta = 1.0$.
In both the Libimseti and synthetic data experiment, the baseline and batch IPFP methods utilized the imputed preference matrix, which is the product of these factor vectors. The mini-batch IPFP directly employed the factor vectors.

We evaluate our method and the baselines in terms of the expected total number of matches, which is calculated by social welfare in a two-sided market, as defined in \cite{Su2022}. 
To simulate user groups with highly congested populations, we used the exponentially decaying examination function in the position-based model.
\begin{equation}
    v(k) = 1 / \exp(k-1),
\end{equation}
, where $k$ denotes the index in the ranking list presented to the candidate or employer.
For synthetic datasets we calculate social welfare using the generated preference matrix.
For the Libimseti dataset, we calculate social welfare using the imputed preference matrix.

\subsubsection{Results.}
\begin{figure}[tbp]
    \includegraphics[width=0.5\textwidth]{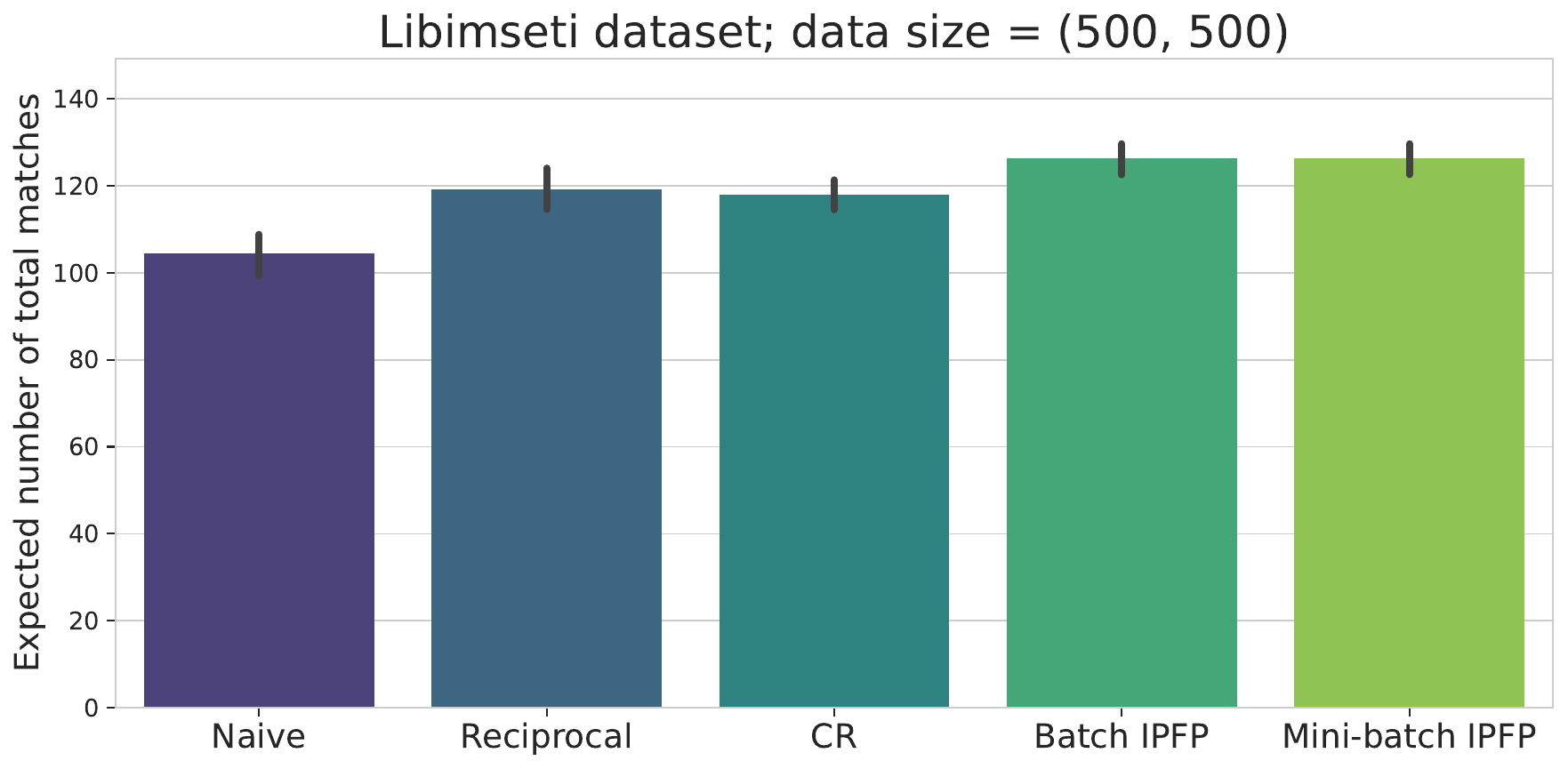}
  \caption{Results of Libimseti data experiments. The market size is 500 men and 500 women, and the examination function is $v(k) = 1 / \exp(k-1)$.}
\label{experiment_sw_realdata}
\end{figure}

\begin{figure*}[tbp]
    \includegraphics[width=0.9\textwidth]{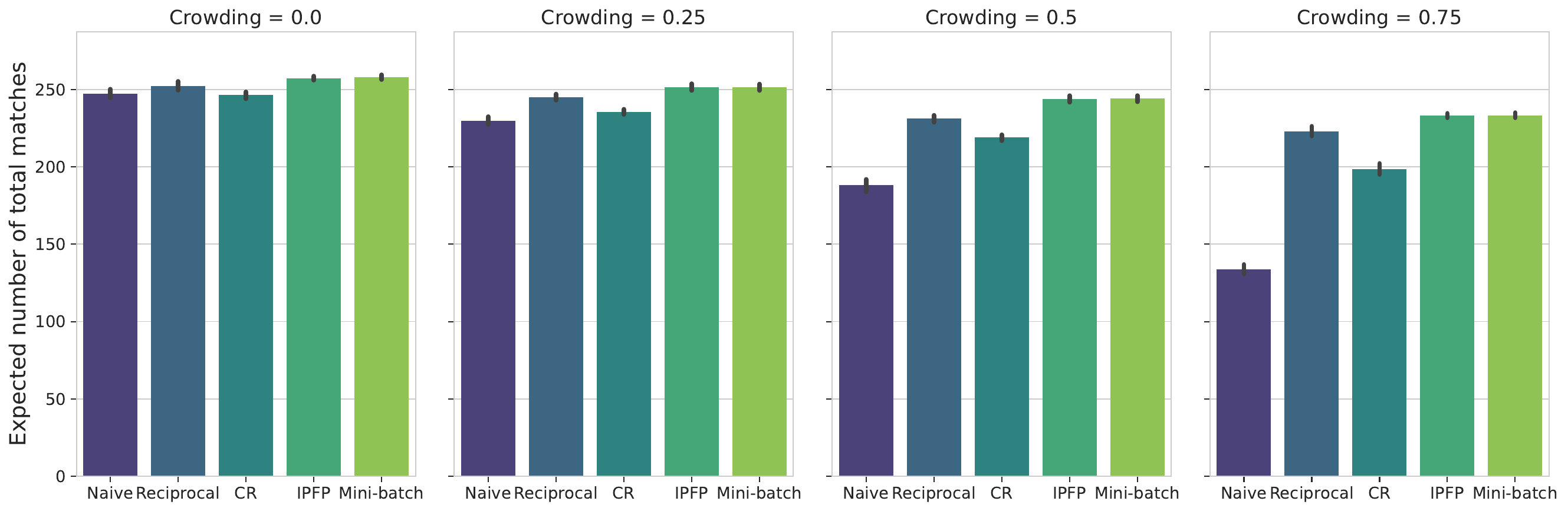}
  \caption{
  Synthetic data experiment results at various crowding parameter levels. The market size is 500 jobs and 1000 candidates, and the examination function is $v(k) = 1 / \exp(k-1)$. When following the IPFP-based recommendation policy, the expected number of matches remains higher than the baseline methods as crowding parameters increase.  For mini-batch IPFP, the number of matches slightly decreases because the preference matrix is approximated using the product of factor vectors.}
\label{experiment_sw_syntheic}
\vspace{-0.5cm}
\end{figure*}

Figure \ref{experiment_sw_realdata} presents the results of the Libimseti dataset.
The batch and mini-batch IPFP methods demonstrated the highest number of matches, indicating its effectiveness in optimizing match outcomes.
Figure \ref{experiment_sw_syntheic} presents the results of the synthetic data experiments conducted with various crowding parameters. 
The results demonstrate that the IPFP-based recommendation policy maintains superior performance as crowding parameters increase. Specifically, the expected number of matches not only remains consistently higher than the baseline methods but also exhibits remarkable resilience to performance degradation under increased crowding conditions.
This suggests that IPFP is particularly effective in highly crowded markets.
Considering that factor vector-based preference matrix calculations, such as matrix factorization, are commonly used for large-scale real-world recommender systems, we believe that mini-batch IPFP remains a viable approach for efficiently handling large-scale matching problems.

\subsection{Experiments on Computational Efficiency}
\subsubsection{Datasets.}
We generated synthetic market data with the same number of candidates and employers.
For batch IPFP experiments, we set the sample size parameters $|\mathcal{X}|$ = $|\mathcal{Y}|$ in $\{10^2, 10^3, 10^4\}$.
For mini-batch IPFP experiments, we further expanded the size parameters in $ \{10^2, ..., 10^6\}$.

To calculate the unidirectional preference score, we sampled the factor vectors of candidates and employers
$\boldsymbol{f}_x, \boldsymbol{k}_x, \boldsymbol{g}_y, \boldsymbol{l}_y$ from a uniform distribution $U\left[0, \frac{1}{\sqrt{D}}\right]$, where the dimension of the factor vector is $D=50$ during the experiments.
We assumed that all users have the same capacity value -- $\forall x \in \mathcal{X}, \:  n_x=C/|\mathcal{X}|$ and $\forall y \in \mathcal{Y}, \:  m_y=C/|\mathcal{Y}|, $ where $C$ is a constant value. 


\subsubsection{Algorithms and Metrics.}
We compared the following four IPFP variants in terms of computational cost and memory efficiency:
(a) Batch IPFP on CPU (vanilla IPFP, as baseline), (b) Batch IPFP on GPU, (c) Mini-batch IPFP on CPU, and (d) mini-batch IPFP on GPU.
We set $\beta = 1.0$ for all methods. 
For mini-batch IPFP methods, we experimented with various batch sizes in $B \in \{1, 10, 100\}$ to fit within our computer memory.
We executed iteration $I=100$ loops and evaluated the average calculation time per loop and the overall CPU or GPU memory consumption.

We also measured the computation time and memory consumption for a population of $|\mathcal{X}|$ = $|\mathcal{Y}| = 10^4$ while varying the dimension of factor vectors in the mini-batch IPFP.
\footnote{We conducted experiments on an Intel Core i9-12900K CPU and single NVIDIA GeForce RTX 3080 (10GB memory) GPU computer.}

The source codes were written by inheriting from OTT-JAX~\cite{cuturi2022optimal}, a solver kit for solving optimal transport problems using the vector computation library JAX~\cite{jax2018github}. \footnote{https://github.com/74hcnklULDuids89/minibatch-ipfp.git}

\subsubsection{Results.}

\begin{figure}[tbp]
  \begin{minipage}[b]{\columnwidth}
    \centering
    \includegraphics[keepaspectratio, width=\columnwidth]{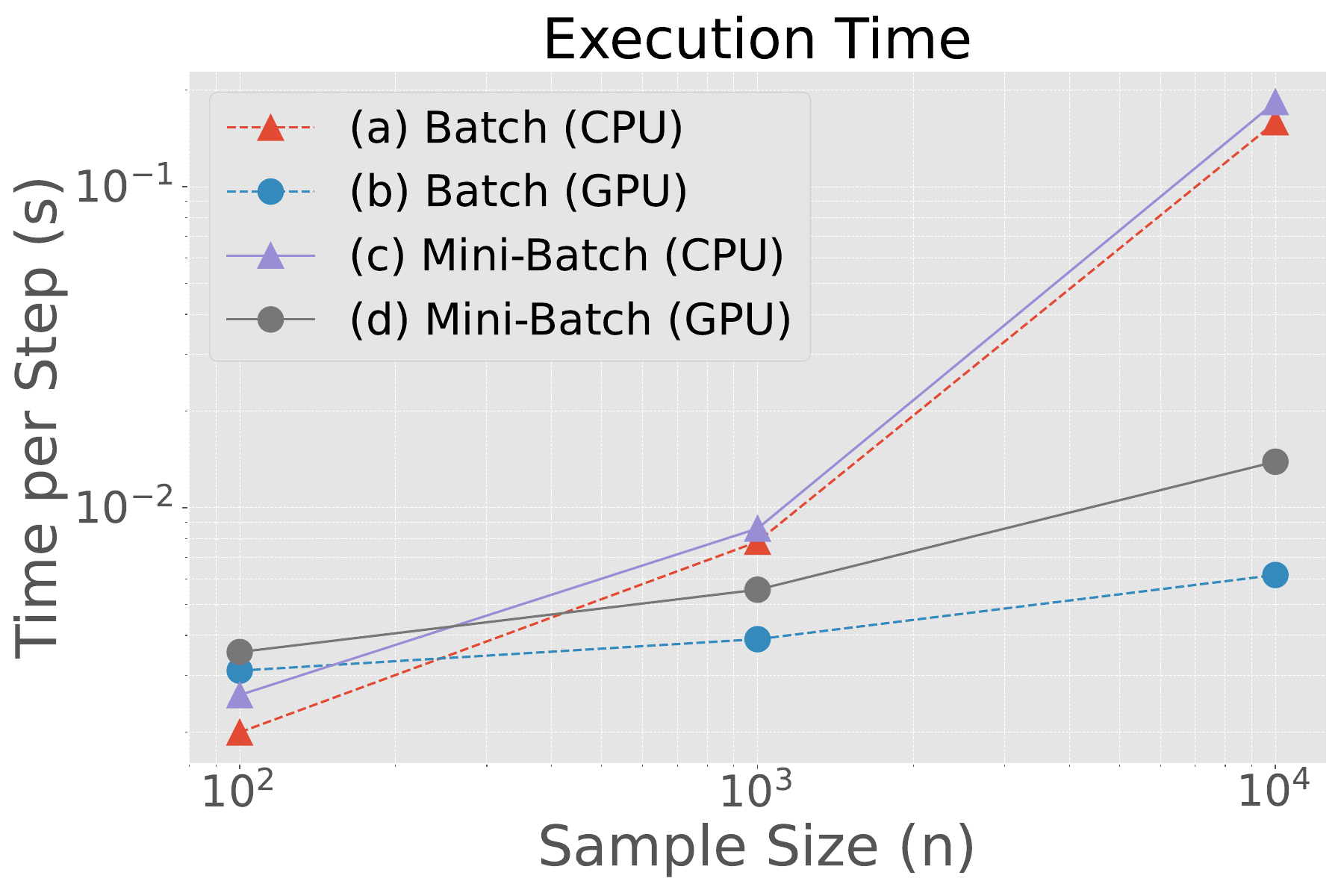}
  \end{minipage}
  \begin{minipage}[b]{\columnwidth}
    \centering
    \includegraphics[keepaspectratio, width=\columnwidth]{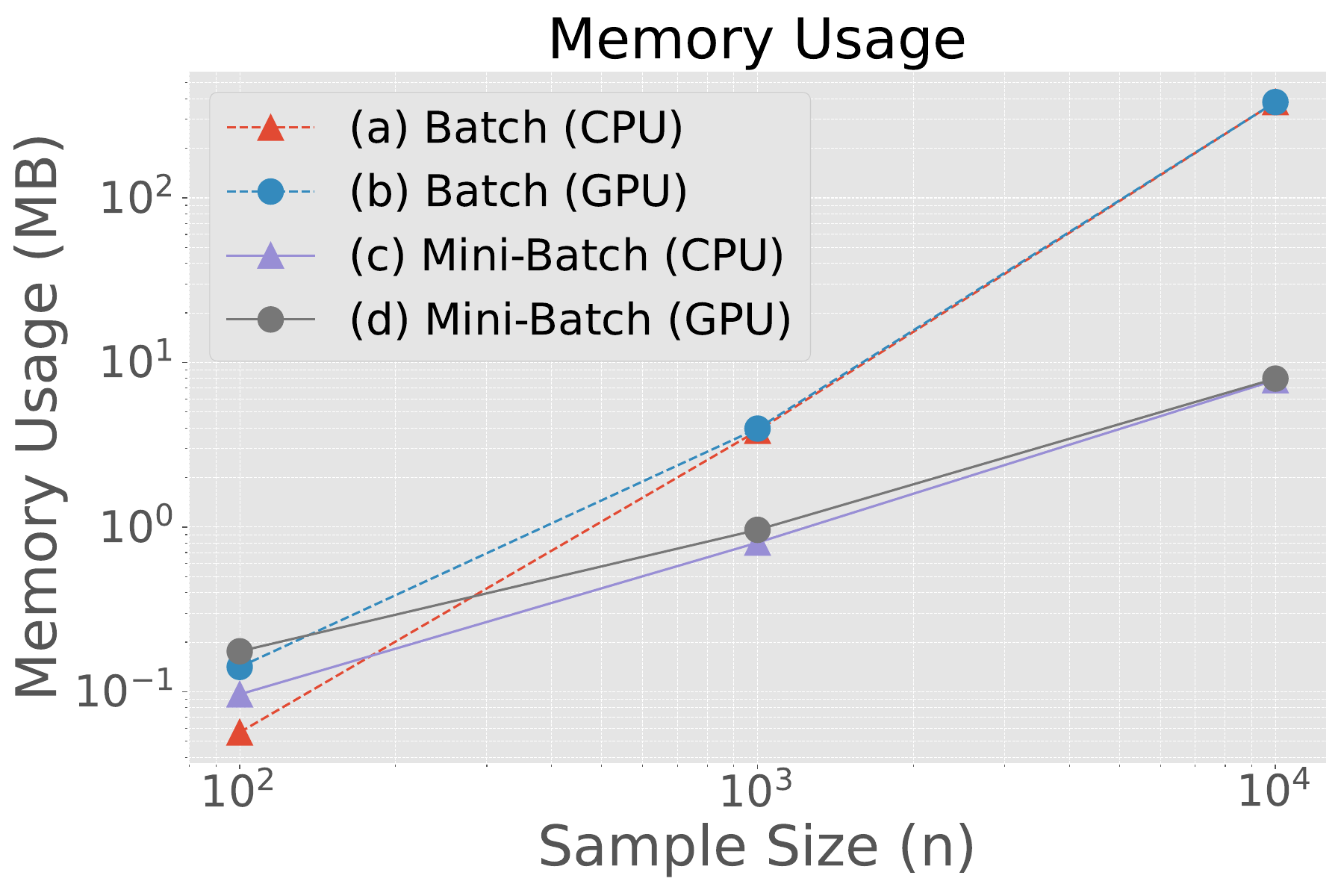}
  \end{minipage}
  \caption{Calculation time (left) and memory usage (right) of batch and mini-batch IPFP with varying data size values. The average calculation times span over 100 iterations. Memory usage is measured on CPU memory for (a) and (c) and GPU memory for (b) and (d).}
\label{fig_BatchMinibatch}
\end{figure}

Figure \ref{fig_BatchMinibatch} presents the average computation time per iteration for each method. 
The GPU implementation of IPFP is faster than the CPU implementation for both batch and mini-batch. 
Batch IPFP requires more memory than mini-batch IPFP because it requires loading the preference matrices into memory in advance. 
In our experimental environment, an out-of-memory error prevented execution when the data size exceeded $10^5$.
The mini-batch IPFP handled memory consumption via its online factor vector product calculation and yielded a calculation time comparable to batch IPFP.

\begin{figure}[tbp]
  \begin{minipage}[b]{\columnwidth}
    \centering
    \includegraphics[keepaspectratio, width=\columnwidth]{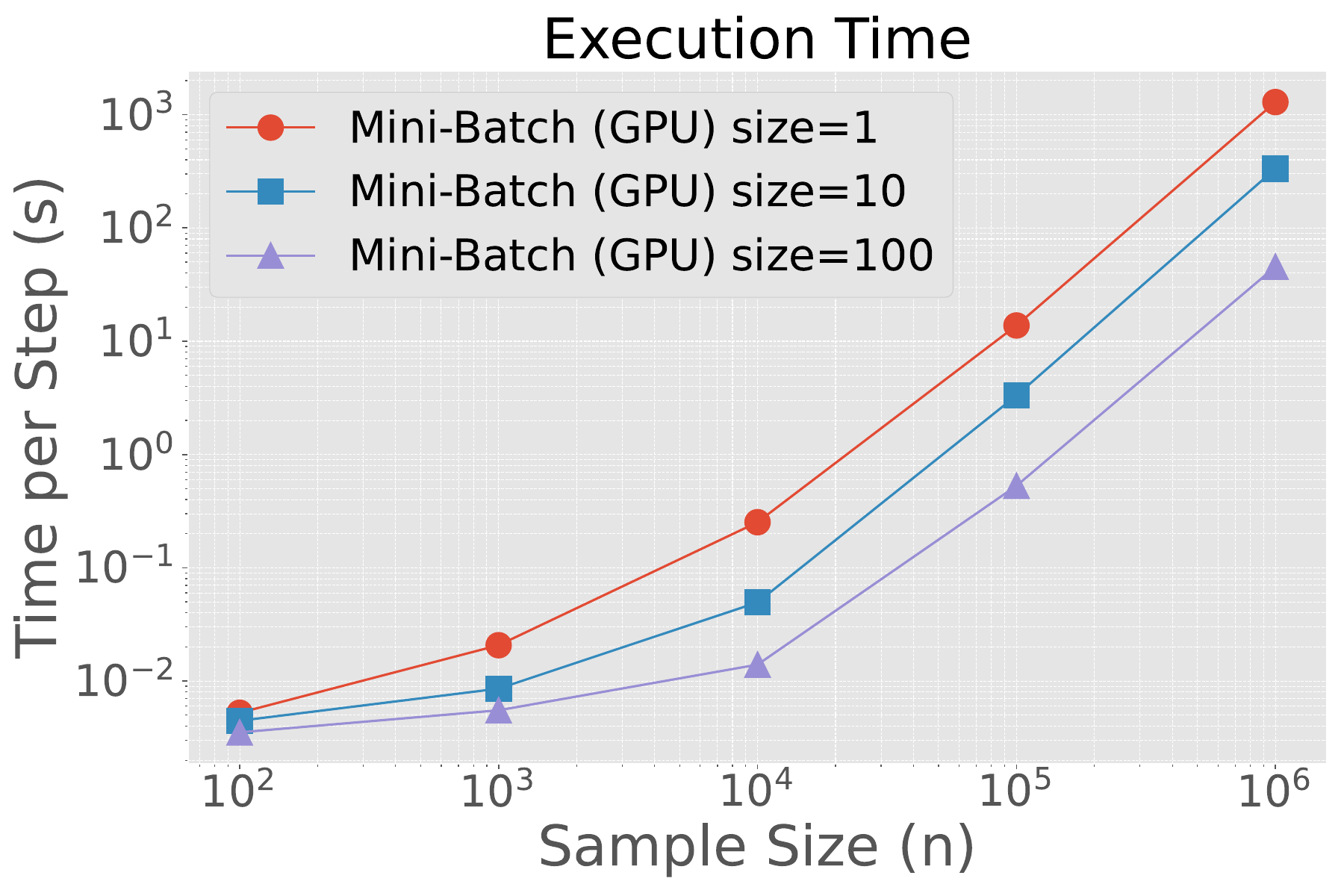}
  \end{minipage}
  \begin{minipage}[b]{\columnwidth}
    \centering
    \includegraphics[keepaspectratio, width=\columnwidth]{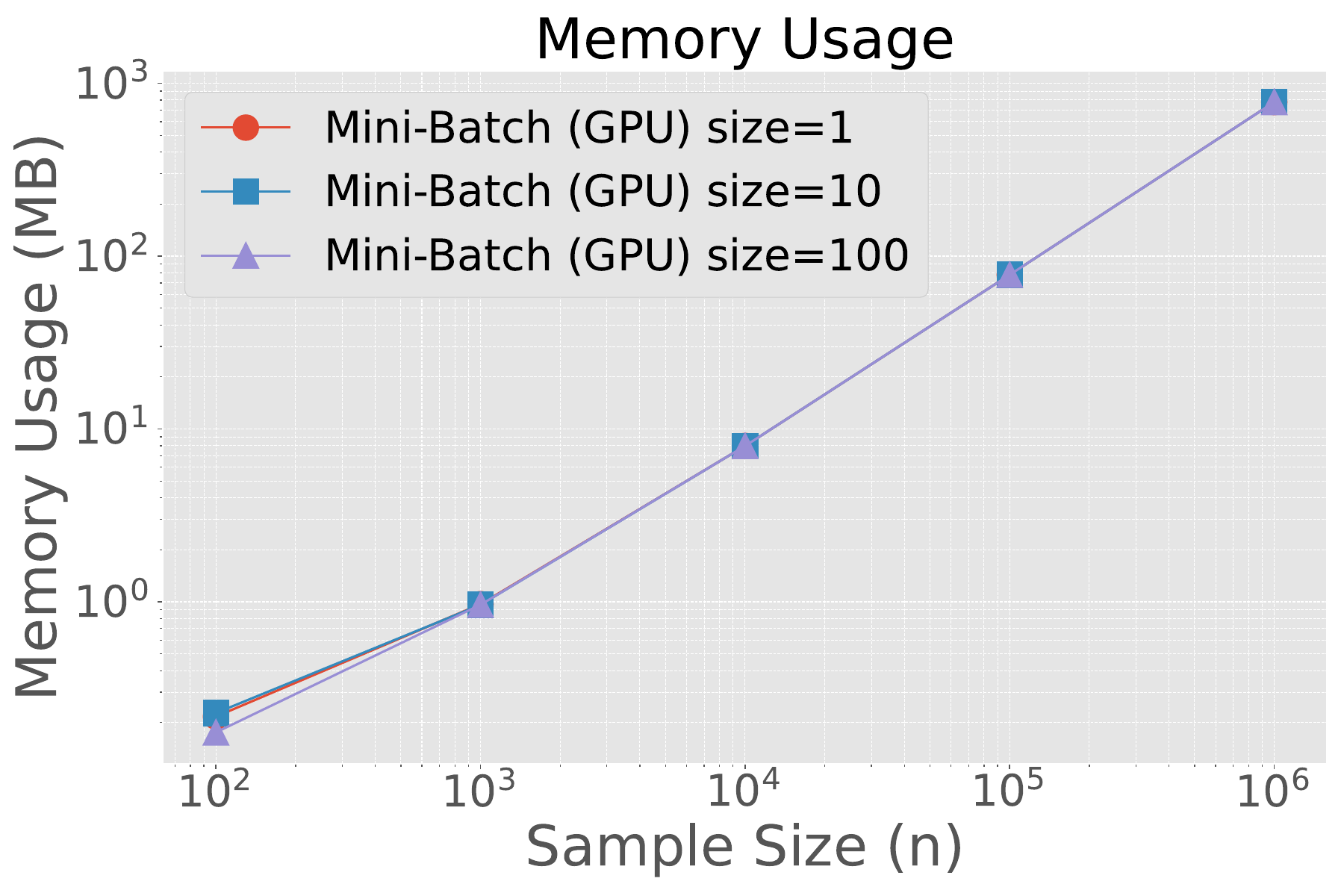}
  \end{minipage}
  \caption{Calculation time and memory usage of mini-batch IPFP for large sample sizes with various batch sizes. The average calculation times span over 100 iterations.}
\label{fig_Minibatch}
\vspace{-0.5cm}
\end{figure}
Figure \ref{fig_Minibatch} presents the computation time and memory usage of mini-batch IPFP for large datasets with different batch sizes.
The execution time increases by a constant factor with the batch size. 
However, because of the effective memory processing of JAX, memory usage remains linear scaling regardless of batch size.
In our calculation environment, setting $B=100$ allows IPFP to be performed in tractable time even with a data size of $10^6$.
In practical applications to real-world scenarios, it is conceivable that the overall computation time could be reduced by implementing an early stopping criterion. This could be achieved by terminating the process when the ranking for candidates and employers remains stable for a predetermined number of epochs.

\begin{figure}[tbp]
  \begin{minipage}[b]{\columnwidth}
    \centering
    \includegraphics[keepaspectratio, width=\columnwidth]{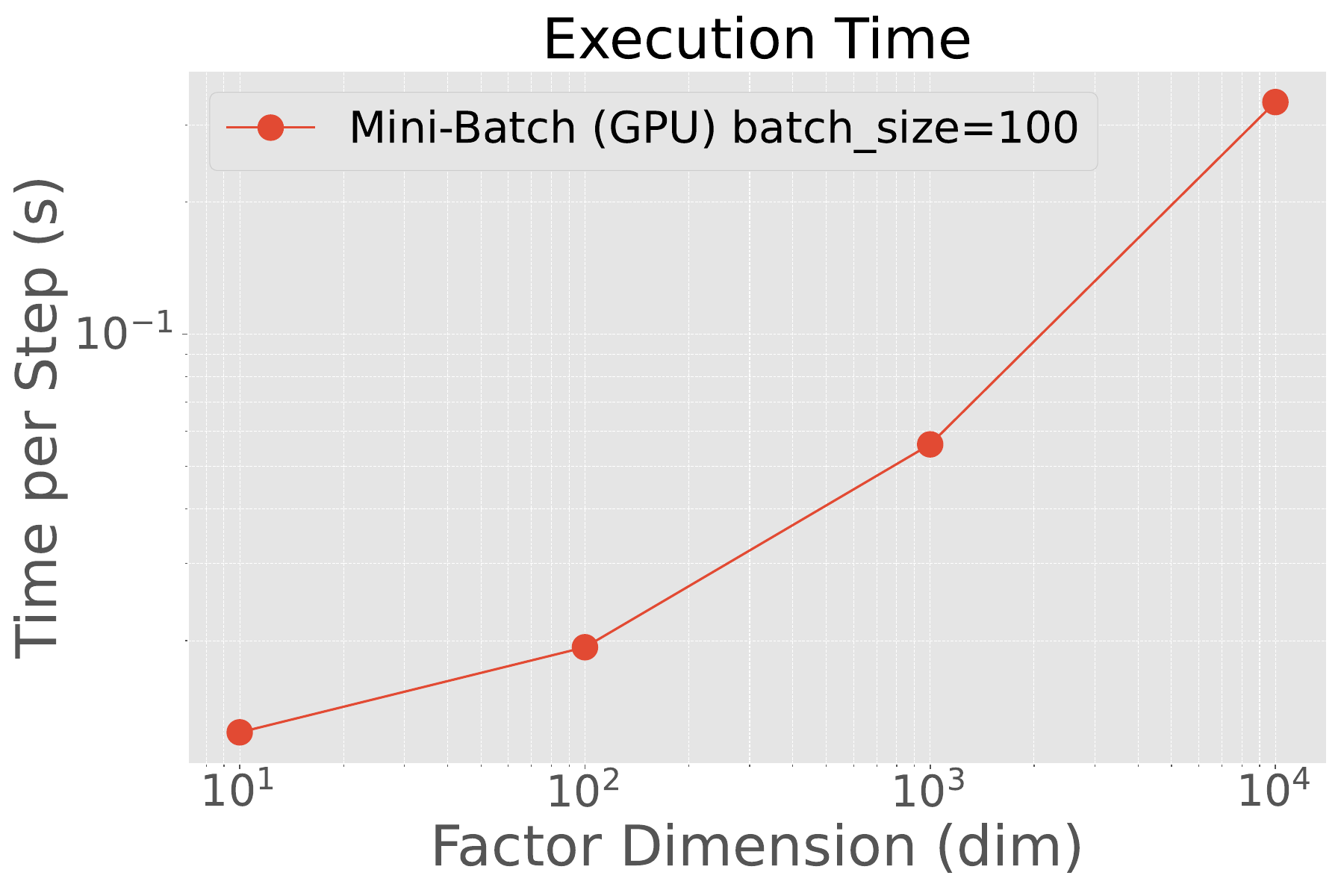}
  \end{minipage}
  \begin{minipage}[b]{\columnwidth}
    \centering
    \includegraphics[keepaspectratio, width=\columnwidth]{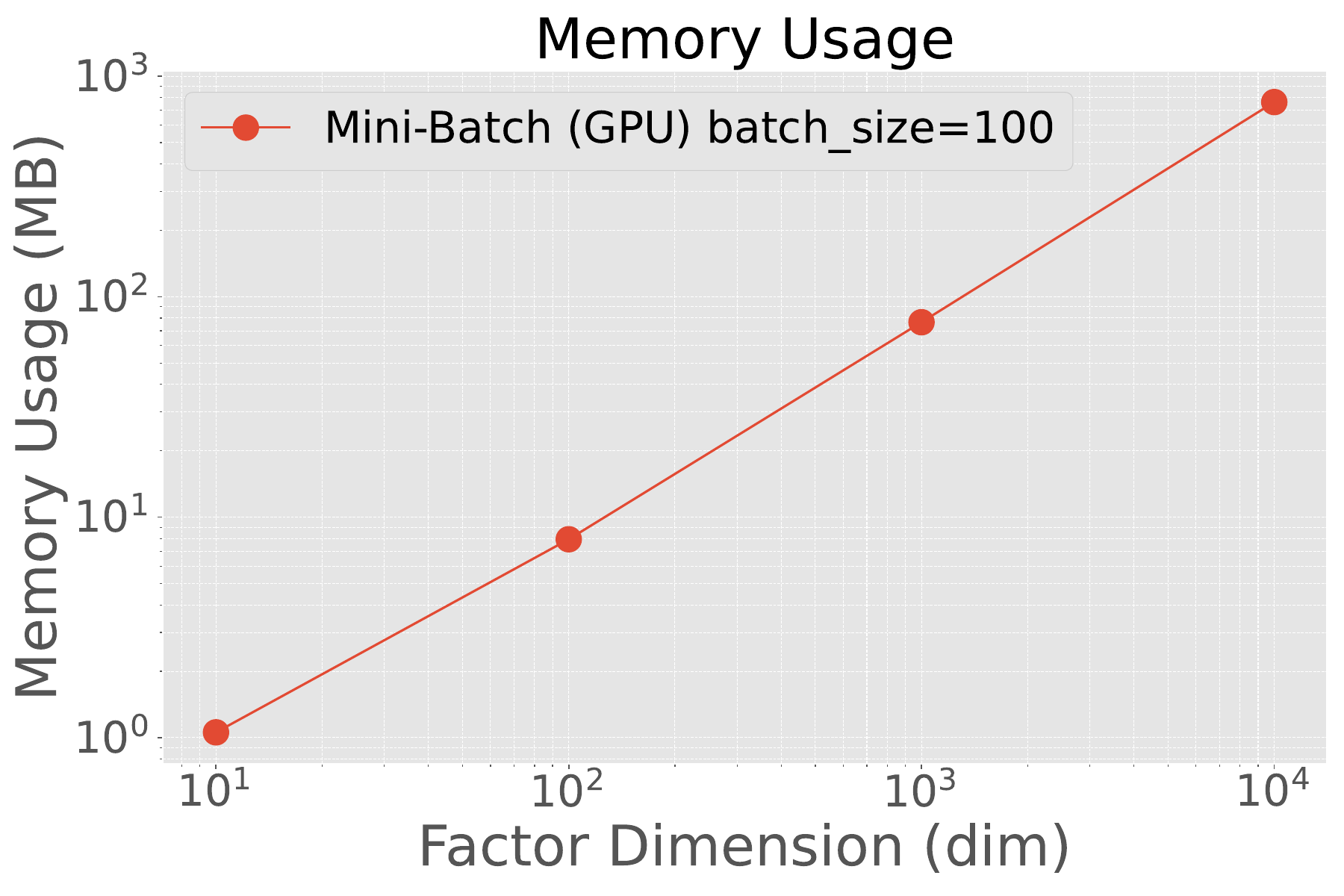}
  \end{minipage}
  \caption{Calculation time and memory usage of mini-batch IPFP for various dimensions of factor vectors. The average calculation times span over 100 iterations.}
\label{fig_dim}
\vspace{-0.5cm}
\end{figure}

Figure \ref{fig_dim} presents the results on the computation time and memory efficiency of the algorithm on synthetic dataset with various dimensions of factor vectors. The increase in computation time and memory consumption of minibatch-IPFP has an almost linear relationship with the increase in the dimension of factor vectors.


\section{Conclusion and Future Work}
In this study, we propose a novel method that significantly improves the computational efficiency and memory usage of the IPFP algorithm to use the TU matching theory in RSSs with large data. Our method achieved the same matching probabilities as the conventional IPFP algorithm, while operating efficiently even for large-scale data using parallel and mini-batch computation techniques.
Our experiments on synthetic and real datasets have demonstrated that our method can process computations for up to a million users, even in typical CPUs/GPUs. 


In future work, we plan to apply the acceleration techniques developed in the field of OT problems. 
Initially, we approximated the transportation cost matrix using the low-rank Sinkhorn factorization algorithm\cite{Meyer2021LRSink}, reducing the spatial complexity.
Second, we calculated the derivative values for the preference matrix and backpropagated them to a unilateral recommendation model~\cite{Yoosof2023Recon}. 

\bibliographystyle{plain}
\bibliography{refs}

\newpage


\appendix
\section{Derivation of Transferable Utility Matching}
\label{Appendix_tu}

Let $\mu_{x,y}$ be a probability distribution that specifies a match between candidate $x$ and employer $y$.
The set of feasible matching $\mathcal{M}$ is defined under the following constraint:
\begin{equation}
\begin{aligned}
\label{matching_set}
\mathcal{M}(m, n)=\Bigg\{\mu_{x,y} \geq 0 \mid \sum_{y \in \mathcal{Y}_0} \mu_{x,y} &= m_x \: \forall x \in \mathcal{X}, \\ 
\sum_{x \in \mathcal{X}_0} \mu_{x,y} &= n_y \: \forall y \in \mathcal{Y} \Bigg\},
\end{aligned}
\end{equation}
where $m_x$ and $n_y$ are the normalized masses of the candidate group $x$ and employer group $y$, respectively. 
We set the total mass of each group to a constant $C$.

\begin{equation}
\label{totalmassconst}
\sum_{x \in \mathcal{X}} n_{x} = \sum_{y \in \mathcal{Y}} m_{y} = C .
\end{equation}

We assume that the two random utility distributions $\mathcal{P}$ and $\mathcal{Q}$ are independent and identically distributed (i.i.d.) with a type-I extreme value distribution with a scale parameter $\beta > 0$. 
Equation (\ref{matching_set}) is transformed into the following equation:

\begin{equation}
\label{Gumbel assumption}
\begin{aligned}
\mu_{x,y}&=m_x \frac{\exp \left(p_{x,y}+\tau_{x,y}\right)}{\sum_{y^{\prime} \in \mathcal{Y}_0} \exp \left(p_{x,y^{\prime}}+\tau_{x,y^{\prime}}\right)} \\
    &=n_y \frac{\exp \left(q_{y,x}-\tau_{x,y}\right)}{\sum_{x^{\prime} \in \mathcal{X}_0} \exp \left(q_{y,x^{\prime}}-\tau_{x^{\prime},y}\right)} .
\end{aligned}
\end{equation}

Let $\phi_{x,y} = p_{x,y} + q_{y,x}$ be an observable joint utility.
According to ~\cite{Galichon2021}, the optimization problem in $\tau_{x, y}$ in (\ref{Gumbel assumption}) 
is a dual expression of the convex (primary) optimization problem for social welfare $W$.

\section{Algorithm of Batch and Mini-Batch IPFP}
\label{Appendix_alg}
Algorithm \ref{alg1} displays the algorithm used in the batch IPFP computation.
Algorithm \ref{alg2} displays the algorithm used in the computation of the mini-batch IPFP.

\begin{algorithm}[tbp]
    \caption{Solving Equilibrium Matching by \textit{Batch} IPFP}
    \label{alg1}
    \begin{algorithmic}[1]
    \REQUIRE preference score matrices $\boldsymbol{P}, \boldsymbol{Q}$ in size $(|\mathcal{X}|, |\mathcal{Y}|)$, normalized mass vectors $\boldsymbol{m}$ of size $|\mathcal{X}|$, $\boldsymbol{n}$ of size $|\mathcal{Y}|$, scale parameter $\beta$
    \REQUIRE a maximum number of iterations $I$
    \ENSURE a matrix $\boldsymbol{\mu}$ of size $(|\mathcal{X}|, |\mathcal{Y}|)$ denotes an equilibrium matching pattern.
    \STATE $\boldsymbol{u} \leftarrow \boldsymbol{1}$ \COMMENT{size ($|\mathcal{X}|$)}
    \STATE $\boldsymbol{v} \leftarrow \boldsymbol{1}$ \COMMENT{size ($|\mathcal{Y}|$)}
    \STATE $\boldsymbol{A} \leftarrow \exp(\frac{\boldsymbol{P} + \boldsymbol{Q}}{2\beta}) $ 
    
    \FOR{$ i = 1, ..., I$} 
        \STATE $\boldsymbol{s} \leftarrow \boldsymbol{A} \boldsymbol{v} / 2 $
        \STATE $\boldsymbol{u} \leftarrow \sqrt{\boldsymbol{s}^2+ \boldsymbol{m}}-\boldsymbol{s} $
        \STATE $\boldsymbol{s} \leftarrow \boldsymbol{A}^{T} \boldsymbol{u} / 2$
        \STATE $\boldsymbol{v} \leftarrow \sqrt{\boldsymbol{s}^2+ \boldsymbol{n}}-\boldsymbol{s} $
        \STATE $i \leftarrow i+1$
    \ENDFOR
    \STATE $\boldsymbol{\mu} \leftarrow \boldsymbol{A} \odot(\boldsymbol{u} \otimes \boldsymbol{v}) $ 
    \RETURN $\boldsymbol{\mu}$
    \end{algorithmic}
\end{algorithm}

\begin{algorithm}[tbp]
    \caption{Solving Equilibrium Matching by \textit{Mini-batch} IPFP}
    \label{alg2}
    \begin{algorithmic}[1]
    \REQUIRE preference factor matrix $\boldsymbol{F}, \boldsymbol{K},$ in size $(|\mathcal{X}|, D)$, $\boldsymbol{G}, \boldsymbol{L}$ in size $(|\mathcal{Y}|, D)$, normalized mass vectors $\boldsymbol{m}$ of size $|\mathcal{X}|$, $\boldsymbol{n}$ of size $|\mathcal{Y}|$, scale parameter $\beta$, number of mini-batches $J_x, J_y$
    \REQUIRE a maximum number of iterations $I$
    \ENSURE stable factor matrices $\boldsymbol{\Psi}$ of size $(|\mathcal{X}|, 2D + 2)$ and $\boldsymbol{\Xi}$ in size $(|\mathcal{Y}|, 2D + 2)$.
    \STATE $\boldsymbol{u} \leftarrow \boldsymbol{1}$ \COMMENT{size ($|\mathcal{X}|$)}
    \STATE $\boldsymbol{v} \leftarrow \boldsymbol{1}$ \COMMENT{size ($|\mathcal{Y}|$)}
    \FOR{$ i = 1, ..., I$} 
        \FOR{$ j = 1, ..., J_x$} 
            \STATE $\boldsymbol{A}_j \leftarrow \exp\left(\frac{\boldsymbol{F}_{j} \boldsymbol{G}^T + \boldsymbol{K}_{j} \boldsymbol{L}^T}{2\beta}\right) $ \COMMENT{size ($B, |\mathcal{Y}|$)}
            \STATE $\boldsymbol{s}_j \leftarrow \boldsymbol{A}_j \boldsymbol{v} / 2 $
            \STATE $\boldsymbol{u}_j \leftarrow \sqrt{\boldsymbol{m}_j + \boldsymbol{s}_j^2}-\boldsymbol{s}_j $
            \STATE $j \leftarrow j+1$
            
        \ENDFOR
        \FOR{$ j = 1, ..., J_y$}
            \STATE $(\boldsymbol{A}^T)_j \leftarrow \exp\left(\frac{\boldsymbol{F} \boldsymbol{G}^T_j + \boldsymbol{K} \boldsymbol{L}^T_{j} }{2\beta}\right) $ \COMMENT{size ($B, |\mathcal{X}|$)}
            \STATE $\boldsymbol{s}_j \leftarrow (\boldsymbol{A}^T)_j \boldsymbol{u} / 2$
            \STATE $\boldsymbol{v}_j \leftarrow \sqrt{\boldsymbol{n}_j+\boldsymbol{s}_j^2}-\boldsymbol{s}_j $
            \STATE $j \leftarrow j+1$
        \ENDFOR
        \STATE $i \leftarrow i+1$
    \ENDFOR
    \STATE $\boldsymbol{\Psi} \leftarrow \text{Concat} \left(\boldsymbol{F}, \boldsymbol{K}, \beta \log \left(\boldsymbol{u}\right), 1\right) $ 
    \STATE $\boldsymbol{\Xi} \leftarrow \text{Concat} \left(\boldsymbol{G}, \boldsymbol{L}, 1, \beta \log \left(\boldsymbol{v}\right)\right) $ 
    \RETURN $\boldsymbol{\Psi}$, $\boldsymbol{\Xi}$
    \end{algorithmic}
\end{algorithm}

\end{document}